# Fast random number generator based on optical physical unclonable functions

KUN CHEN,[1,2] FENG HUANG,[1,2] PIDONG WANG,[1,2] YONGBIAO WAN,[1,2] DONG LI,[1,2] YAO YAO[1,2,*]

[1] *Microsystem and Terahertz Research Center, China Academy of Engineering Physics, Chengdu 610200, China*
[2] *Institute of Electronic Engineering, China Academy of Engineering Physics, Mianyang 621999, China*
*Corresponding author: yaoyao_mtrc@caep.cn*

**We propose an approach for fast random number generation based on homemade optical physical unclonable functions (PUFs). The optical PUF is illuminated with input laser wavefront of continuous modulation to obtain different speckle patterns. Random numbers are fully extracted from speckle patterns through a simple post-processing algorithm. Our proof-of-principle experiment achieves total random number generation rate of 0.96 Gbit/s with verified randomness, which is far faster than previous optical-PUF-based schemes. Our results demonstrate that the presented random number generator (RNG) proposal has great potential to achieve ultrafast random number generation rate up to several hundreds of Gbit/s.**

Random numbers are essential in many fields, including secure communication [1], gambling industry [2] and numerical simulation [3]. Various random number generators (RNGs) have been developed over the past decades, such as RNGs based on thermal noise [4], radioactive decay [5], laser phase noise [6], quantum vacuum fluctuations [7–10], single photon emitters [11,12], amplified spontaneous emission (ASE) [13], and optical parametric oscillators (OPO) [14]. Among them, optical RNGs provide a promising candidate for random number generation owing to advantages of high speed and resistance to external interference [15]. Recently, RNGs based on physical unclonable functions (PUFs) [16] or especially optical PUFs [17–19] begin to attract the attention of researchers. The optical PUF proposed by Pappu et al. [20] is composed of inhomogeneous materials with randomly distributed micro-nano structures, which is the only category of considered PUF constructions that is true unclonable [21]. The interactions of coherent light with complex 3D micro-nano structures of optical PUFs exhibit a high degree of complexity and thus are unpredictable, providing the foundation for random number generation. Moreover, these complex interactions can be simply recorded in form of speckle patterns that can be processed into unique random binary sequence, and even only a small variation of coherent light would lead to an absolutely different random binary sequence [22]. A typical optical PUF presents in the form of the optical waveguide [17,19]. In Ref. [17], a RNG based on the optical waveguide is performed through modulations of coherent light wavelength, in which each speckle image can produce approximately 8000-bit random bits and a random number generation rate of 0.46 Mbit/s is achieved. However, the generated binary sequences failed to pass the National Institute of Standards and Technology (NIST) suite. Even though the above scheme is later optimized so as to pass all of the tests in the NIST and 1 Mbit random bits could be obtained from each image [19], tedious processing procedure containing multiple steps is required due to the quality of raw data. Therefore, a fast and simple RNG based on optical PUFs is worth exploring.

In this work, we report a fast RNG based on optical PUFs with total random number generation rate up to 0.96 Gbit/s. The input laser wavefront is continuously modulated to illuminate the optical PUF and speckle images are recorded simultaneously. The min-entropy of the speckle images is estimated to calculate the best extraction ratio of randomness extraction. Subsequently SHA-256 hash function [23] is performed to extract random numbers from speckle images. On the premise of ensuring high-quality randomness, the amount of data produced by a single speckle image has been increased up to 28 Mbit. Our RNG has significantly improved upon previous RNGs based on optical PUFs [17–19] in terms of the amount of data from a single speckle image and the total random number generation rate. In addition, the resulting random binary sequence passes all statistical tests of NIST, TestU01, DIEHARD and DIEHARDER suites.

Our optical PUFs are fabricated by Zirconium oxide ($ZrO_2$) nanoparticles with an average grain size of 200 nm, which are randomly immersed in polymethy methacrylate (PMMA) films with a thickness about 1 mm. The schematic diagram of the RNG based on optical PUFs is sketched in Fig. 1. A compact laser diode with central wavelength of 638 nm (Integrated Optics, No. 0638L-11A) is used as the light source. After being expanded, the laser beam is sent to a phase-only liquid crystal spatial light modulator (LCSLM, 1920 × 1080 pixels, pixel size = 8.0 μm, Holoeye, PLUTO-2-VIS-014), on which a phase pattern is displayed to modulate the wavefront of the laser beam. Then the modulated reflected light is projected onto the surface of the PUF through Lens 1. Passing through the PUF, the scattered light will be detected by an industrial CCD camera (Point Gray, CM3-U3-50S5M) with 2448 × 2048 pixels through Lens 2. To obtain raw speckle patterns, a large number of unique pseudo-random phase patterns are sequentially loaded into LCSLM with phase value of each pixel uniformly distributed in the interval [0, 2π]. At the same time, the CCD camera records the speckle pattern corresponding to each LCSLM configuration. Subsequently, a selected post-processing algorithm is applied to the recorded speckle patterns (images) for producing the random binary sequence.

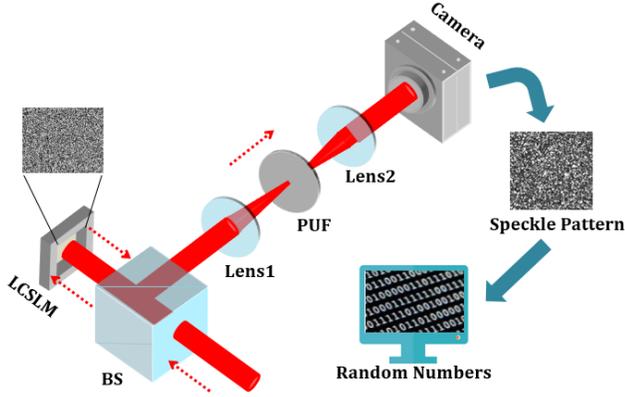

Fig. 1. Schematic diagram of the experimental setup. An expanded laser beam is sent to LCSLM. Then a non-polarizing beam splitter (BS) redirects modulated reflected light toward the PUF, from which scattered light is detected by camera.

A preliminary analysis of raw speckle images using Hamming distance and Euclidean distance metrics is performed to verify the physical foundation of random number generation. Under the same LCSLM configuration, 100 speckle images are collected as intra-dataset to estimate the noise representing the robustness of the system. Another 200 speckle images under the different LCSLM configuration are captured as inter-dataset to evaluate the unpredictability of the system. The Euclidean distances between normalized images are calculated for inter-dataset and intra-dataset. Similarly, the Hamming distances are obtained between images filtered by Gabor hash algorithm [20]. The histograms of Euclidean distances and Hamming distances are depicted in Fig. 2(a) and 2(b) respectively.

As shown in Fig. 2(a), the mean value of intra-dataset distribution ($\mu_{intra} = 44.54 \pm 17.43$) is nearly 1/20 of inter-dataset distribution ($\mu_{inter} = 816.10 \pm 10.09$) and there is no overlap between them, indicating a relatively low system noise. Moreover, it is clear that inter-dataset distribution is concentrated and has no elongated tail. In Fig. 2(b), the mean value for inter-dataset distribution is 0.4987 and the coefficient of variation (ratio of standard deviation to mean) is approximately 0.0036, which show that Hamming distance values are centered on the mean of 0.4987. A mean value very close to 0.5 and a fairly small coefficient of variation also imply a high degree of unpredictability under the different LCSLM configuration, providing the necessary condition for random number generation.

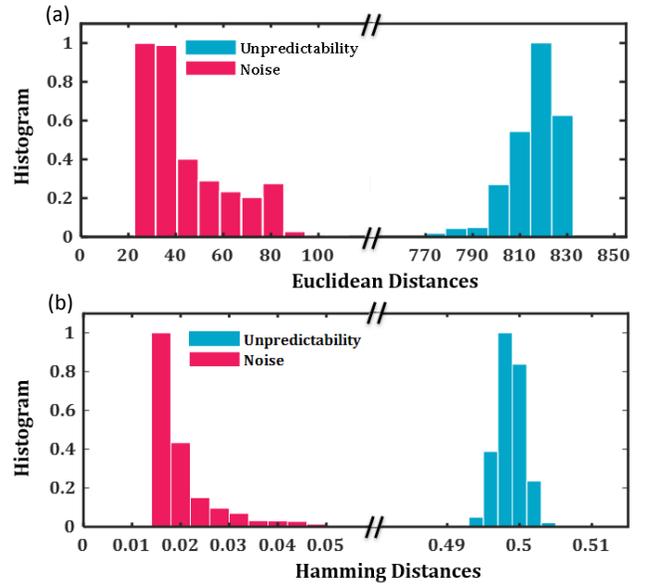

Fig. 2. Histograms of (a) Euclidean distances and (b) Hamming distances for intra-dataset under the same LCSLM configuration and for inter-dataset under the different LCSLM configuration.

The first step in random number generation of raw speckle images is to develop a quantitative understanding of their randomness. The amount of randomness can be quantified with the min-entropy, which represents the conservative lower bound of randomness [24,25]. Herein, we adopt the min-entropy to assess randomness of raw speckle images from inter-dataset and evaluate best extraction ratio for randomness extraction.

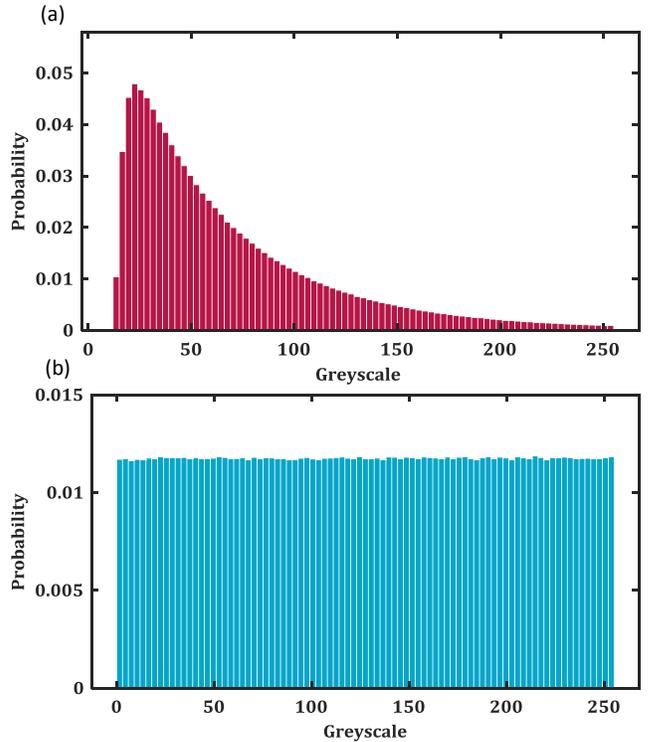

Fig. 3. Probability distribution of the grey value for (a) a raw speckle image and (b) the same image after randomness extraction via SHA-256.

The min-entropy of raw images is calculated to be 5.95, which means that only 5.95 information-theoretically random bits can be generated from 8 raw bits. The ideal min-entropy of 8 is not reached and it reveals a relatively low degree of randomness of raw images. Meanwhile, the grey value histogram of the raw image is visualized in Fig. 3(a), which lists the probability distribution of all pixels for the raw image within the grey value of 0-255. The histogram shows that the grey value distribution is not smooth and uniform, thus the randomness of the raw image is not ideal. Overall, both the actual min-entropy of 5.95 and the nonuniform distribution of the grey value illustrate there is still a great space to improve the randomness of raw images. Therefore, it is imperative to perform randomness extraction with regard to raw images. The quality of raw images can be improved using Secure Hash Algorithm (SHA) such as the SHA-256 hash function applied here. Importantly, there is currently no effective method to attack SHA-256 hash function [26]. The grey value distribution of the hashed image is shown in Fig. 3(b). It can be seen that the grey value is uniformly distributed, revealing the favorable statistical characteristics of hashed grey value. However, uniform distribution of grey value does not represent a high degree of randomness of the hashed image. Based on the analysis of min-entropy, the best extraction ratio is $5.95/8 \approx 0.744$ on the premise of ensuring high-quality randomness. As mentioned previously, the SHA-256 we are using in this work, has a fixed output length of 256 bits. Thus, the input bit-string length should be at least $256 \times 8/5.959 \approx 344$ bits in order to generate random bits with high-quality randomness.

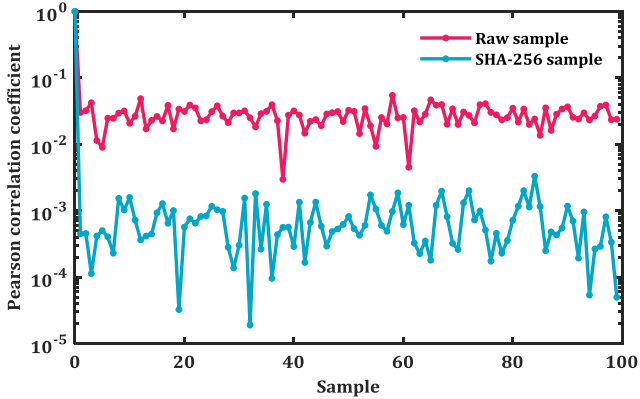

Fig. 4. The Pearson correlation coefficient between 100 different speckle images before and after randomness extraction via SHA-256.

Based on the above analysis, 100 raw speckle images from inter-dataset are utilized to generate the random binary sequence with the best extraction ratio of 0.744. In order to analyze the correlation between speckle images before and after randomness extraction, we have adopted the Pearson correlation coefficient, which is usually used to characterize the correlation between images [27]. As depicted in Fig. 4, hashed results (blue curve) show a relatively low correlation coefficient and a reduction of over two orders of magnitude is observed, implying a higher unpredictability between hashed results. Moreover, the correlation coefficient of hashed results declines to the order of $10^{-4}$, which is sufficiently small to be deemed insignificant. That is, there is nearly ideal independence and unpredictability between hashed results. In addition, we have also analyzed the correlation coefficient between the pseudo-random numbers loaded into LCSLM and the obtained random numbers, and the result ($\sim 10^{-4}$) shows that there is almost no correlation between them.

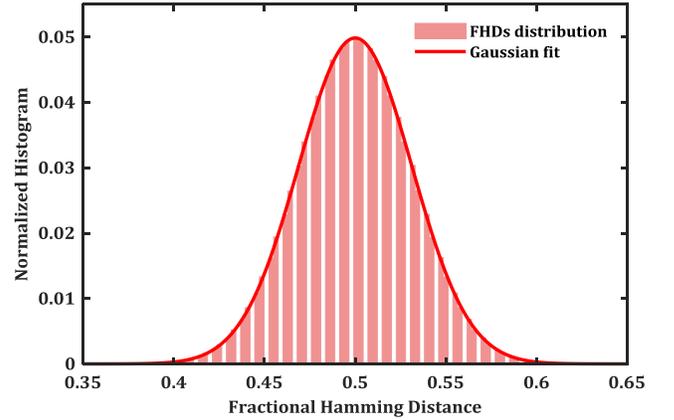

Fig. 5. Hamming distance distribution over 256-bit long strings via SHA-256.

Furthermore, we investigate the distribution of Hamming distances over a large number of 256-bit long strings constituting the random binary sequence to assess the randomness. As an illustration, the Hamming distance distribution obtained from 12.5 million comparisons between 5000 long strings compared with each other are seen in Fig. 5. The mean value μ of Hamming distances is found to be 0.50001 with a standard deviation σ of 0.03125. What is apparent is that the red solid line shows nearly perfect Gaussian curve fitting to the distribution. Using the following formula [20], the degrees-of-freedom (N) of long strings can be estimated:

$$N = \frac{\mu \times (1-\mu)}{\sigma^2} = \frac{0.50001 \times (1-0.50001)}{(0.03125)^2} \approx 256. \quad (1)$$

The degrees-of-freedom and the length of each long string are identical, indicating that these long strings are adequately random and full entropy has been thoroughly extracted from raw speckle images. Collectively, these long strings with full entropy and the nearly perfect Gaussian curve illustrate a good level of randomness of the random binary sequence produced by SHA-256 hash function.

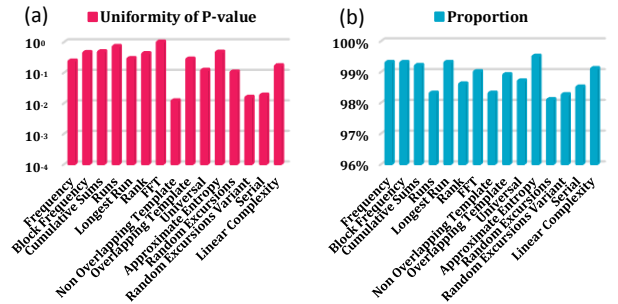

Fig. 6. NIST test results: (a) uniformity of P-value and (b) passed proportion of 1 Gbit random sequence. The minimum of test outcomes is selected for each test item with multiple P-values and proportions.

Finally, we apply the standard test suite NIST SP800-22 [28] consisting of fifteen items to stringently identify the randomness of generated random numbers. In our scenario, each raw speckle image is capable of generating approximately 28 Mbit random bits. Under a typical camera frame rate of 35 fps, there is a random number generation rate of 0.96 Gbit/s. Herein, we adopt 1 Gbit random sequence from 36 successive raw speckle images to perform the NIST test suite, which is read and divided into 1000 subsequences of 1 Mbit. The significance level α is set to 0.01 as recommended by NIST. All the final P-values exceed the significance level of 0.01 and results of NIST statistical test are visualized in Fig. 6. It is obvious that uniformity of P-values is all larger than 0.0001 and proportions passing the test are all above the threshold of 98%, indicating the 1 Gbit random sequence successfully pass the NIST test suite. We also apply the TestU01 Alphabit battery, DIEHARD and DIEHARDER suite to test the 1 Gbit random sequence. The results show that our data have passed all the statistical tests successfully in TestU01 Alphabit battery and DIEHARD suite. For the more stringent DIEHARDER suite, our random sequence passes all the tests with only four "weak" results.

In conclusion, we develop a fast and simple RNG with our homemade optical PUFs. The input laser wavefront is continuously modulated by LCSLM to illuminate the optical PUF, from which scattered light is recorded in the form of speckle patterns. The raw speckle patterns without any pre-processing procedures are used to directly generate random numbers by SHA-256 hash function. The amount of data produced by a single speckle image is 28 Mbit on the premise of ensuring high-quality randomness, which is more than 20 times higher than other RNGs based on optical PUFs [19]. It is worth mentioning that the current random number generation rate with verified randomness is 0.96 Gbit/s in our proof-of-principle experiment, which is far faster than the ever reported 0.46 Mbit/s [17]. The improved rate can be mainly attributed to following two aspects: (1) the homemade optical PUF is designed to be more complex in structure than optical waveguide used in Refs. [17-19] with finite transverse modes, where we have controlled the grain size of nanoparticles, thickness of PUF and number of nanoparticles appropriately; (2) the selected post-processing procedure is well suitable for random number generation, where we have taken full advantage of the greyscale information present in the raw speckle images. In near future, random number generation rate can be lifted to several hundreds of Gbit/s if we adopt hardware devices with superior performance in our proposal. For example, we can implement a high-speed digital micromirror device (Vialux-V-7001) and a comparable camera (CP70-004-M/C-19000).

**Funding.** Science Challenge Project (TZ2018003); National Natural Science Foundation of China (61875178, 12175204, 61805218, 12104423).

**Disclosures.** The authors declare that there are no conflicts of interest related to this article.

**Data Availability.** Data underlying the results presented in this Letter are not publicly available at this time but may be obtained from the authors upon reasonable request.